**Michał STANISZEWSKI**, Andrzej POLAŃSKI
INSTITUTE OF INFORMATICS, SILESIAN UNIVERSITY OF TECHNOLOGY,
Akademicka Street 16, 44-100 Gliwice, Poland

# Hankel Singular Value Decomposition as a method of preprocessing the Magnetic Resonance Spectroscopy

**Mgr inż. Michał STANISZEWSKI**

PhD student in Informatics since 2010 at Silesian University of Technology in Gliwice. Beneficiary of Scholarship DoktoRIS -program for innovative Silesia co-financed by the European Social Fund. Cooperation with Cancer Center and Institute of Oncology Gliwice Branch as an IT specialist. Research area: magnetic resonance spectroscopy and imaging, computer graphics.

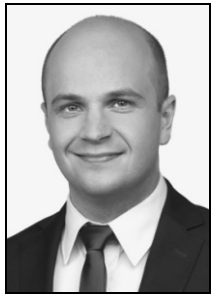

*e-mail: Michal.Staniszewski@polsl.pl*

**Prof. dr hab. inż. Andrzej POLAŃSKI**

Professor in Institute of Informatics in Silesian University of Technology and in Polish-Japanese Institute of Information Technology. Research Interests in Bioinformatics and Biometrics: genetics statistical modeling of evolution, the theory of coalescence, comparative genomics, analysis of genomic sequences, DNA sequencing, high-pass data of molecular biology, analysis of microarray data, design of classifiers based on microarray data.

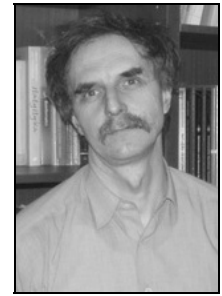

*e-mail: Andrzej.Polanski@polsl.pl*

## Abstract

The signal resulting from magnetic resonance spectroscopy is occupied by noises and irregularities so in the further analysis preprocessing techniques have to be introduced. The main idea of the paper is to develop a model of a signal as a sum of harmonics and to find its parameters. Such an approach is based on singular value decomposition applied to the data arranged in the Hankel matrix (HSVD) and can be used in each step of preprocessing techniques. For that purpose a method has was tested on real phantom data.

## Rozkład macierzy Hankela według wartości osobliwych jako metoda do przetwarzania wstępnego spektroskopii rezonansu magnetycznego

### Streszczenie

Sygnał pochodzący z badania spektroskopii rezonansu magnetycznego zawiera również liczne szumy oraz nieprawidłości, stąd aby zastosować wyniki jako narzędzie diagnostyczne należy wprowadzić kilka usprawnień. W tym celu stosuje się filtrowanie, korekcję linii bazowej, korekcję fazy, korekcję prądów wirowych oraz usuwanie niechcianych komponentów, które nazywa się przetwarzaniem wstępnym. W dalszej analizie bardzo ważna jest identyfikacja poszczególnych metabolitów, którą można otrzymać poprzez zamodelowanie sygnału. Głównym pomysłem przedstawionym w artykule jest rozwinięcie modelu sygnału jako sumy harmonicznej. Metoda polega na znalezieniu parametrów opisujących sygnał takich jak amplituda, przesunięcie fazowe, częstotliwości i współczynnik tłumienia. Takie podejście bazuje na rozkładzie macierzy według wartości osobliwych (SVD) zastosowanym na danych zawartych w macierzy Hankela (HSVD), który dekomponuje sygnał na sumę harmonicznych oraz wylicza potrzebne parametry. Autor zaproponował zastosowanie HSVD w technikach przetwarzania wstępnego. Artykuł opisuje główne kroki przetwarzania i rozwiązanie każdej części oparte na HSVD. Podsumowując można stwierdzić, iż HSVD stosuje się w dekompozycji sygnału ale może być również skutecznym narzędziem w przetwarzaniu wstępnym. Artykuł składa się z 6 rozdziałów, w tym wstępu, rozdziału opisującego HSVD, metody przetwarzania wstępnego i główne wyniki, wniosków i referencji. W artykule znajdują się 4 obrazki oraz 7 referencji.

**Słowa kluczowe:** HSVD, macierz Hankela, rozkład macierzy według wartości osobliwych, przetwarzanie wstępne sygnału MRS.

## 1. Wstęp

The Magnetic Resonance Spectroscopy (MRS) is commonly used as a research technique in current chemistry and medicine. The Nuclear Magnetic Resonance (NMR), which is a physical background for the MRS, is an effect relying on magnetic properties of atomic nuclei. Magnetic nuclei stated in an external magnetic field, due to fast changes of the field, can absorb and register electromagnetic energy. The magnetic resonance is often applied to atomic nuclei of hydrogen, which are present in molecules of water. Almost all human tissues consist of water, however in different proportion in relation to other compounds. Such properties give a chance to register the changes of a resonance emission signal coming from hydrogen which are present in water and tissues. The NMR is a base for two diagnostic methods – Magnetic Resonance Imaging (MRI) and Magnetic Resonance Spectroscopy (MRS). The first method – MRI – gives detailed visualization of internal structures, which in medicine can be used to distinguish a pathologic tissue from the normal one. While the Imaging provides information of the structure, MRS tells more about the biochemical composition. The method is a non-invasive technique which provides the details of metabolic changes in selected tissues. The main difference between the two methods is that the MRI can give only localization of a cancer in a human body, whereas the MRS tells how aggressive the cancer is. The receiver signal detected in the receiving coil of a modern MR spectrometer is known as free induction decay (FID) and it contains all of the resonant frequencies of the sample nuclei. FID represents data in the time domain but to make the results more readable an alternative lookout at the signal was introduced. Instead of reading FID presented in the time evolution, a plot called a spectrum can be drawn. It presents particular components of the signal as a function of frequency. Conversion of time-domain FID into a spectrum is performed by a mathematical method called the Fourier transform (FT) [1].

Due to a number of different limitations, MRS data and its quality is often insufficient. In order to use the results of MRS as a diagnostic tool, several methodological developments are necessary. A low signal to noise coefficient, caused by several limitations of the measurement technique, such as the low concentration of metabolites and the limited time of observation, makes extraction of useful data from MRS signal more difficult. Therefore appropriate preprocessing techniques are used for signal enhancement after the measurement. Most methods are applied on hardware side of an MRS spectrometer, however very often computer software is designed to quantify the MR signal. In order to maximize the information available in the MRS signal, processing techniques can be applied before and after Fourier transformation, since FT operates in both directions, which means that the spectrum can be transformed back to FID without any loss of information [2].

The main concept presented in the paper is based on developing a model as a sum of harmonics. The method consists in finding parameters describing the model, namely: the amplitude, the phase related to the position of the magnetization vector at the beginning, the frequency of oscillations and the damping factor with which the signal decays exponentially. Such an approach is made thanks to the procedures based on singular value decomposition (SVD) applied to the data arranged in the Hankel matrix (HSVD), which decomposes it into a sum of exponentials and in the next step gets all needed parameters. The author proposed an application of the method HSVD in terms of preprocessing techniques. Most of the methods presented in this paper are applied to FID and their result is transformed by FT into a spectrum. It should be noted that different preprocessing methods may be required depending on different reasons. It does not mean that signal enhancement by the use of all methods has to be performed each time.

## 2. Hankel Singular Value Decomposition

MRS data included in the time-domain FID can be described as sum of exponentially damped sinusoids. Such model consists of harmonic components which correspond to MRS peaks in the related frequency domain after FT. Each component of sum is built with amplitude, damping factor, angular frequency and phase according to equation (1). The purpose of method attached in that paper rely on finding those parameters [3].

$$s_n = \sum_{k=1}^{K} a_k e^{\left(-d_k + j\,2\,\Pi f_k\right)n} + j\Phi_k \tag{1}$$

where: $K$ - the number of harmonic components, $a_k$ - amplitude of the $k$-th component, $d_k$ - damping factor of the $k$-th component, $f_k$ - frequency of the $k$-th component, $\Phi_k$ - phase of the $k$-th component, $t$ - sampling step.

The methods of modeling MRS presented in the paper are based on Singular Value Decomposition. SVD gives opportunity to decompose each matrix in the real or the complex form as a multiplication of specific values (2). This decomposition can be applied in signal processing and statistics.

$$M = U \Sigma V^H \tag{2}$$

where: $U$ - unitary matrix of size $m \times m$, $\Sigma$ – diagonal matrix of size $m \times n$ with nonnegative diagonal, $V^H$ - $n \times n$ unitary matrix created as the conjugate transpose of $V$.

The modeling of MRS and removal of any signal present in the spectrum can be performed in the time-domain by the state-space method called HSVD. Such an approach allows the data analysis with application of a more sophisticated algorithm in nonlinear least squares sense. The main drawback of the HSVD is associated with the computational load coming from Singular Value Decomposition used for estimation of the signal subspace.

The data are modeled according to equation (1). For the sake of simplicity at the beginning the assumption that data are noiseless should be stated. HSVD starts with arranging the data in the form of an $L \times M$ matrix. In that case two specific matrices can be used – Hankel matrix $S_H$ where all elements in each antidiagonal are equal (3) or Toeplitz matrix $S_T$ in which each descending diagonal from left to right is equal.

$$S_H = \begin{bmatrix} s_0 & s_1 & s_2 & \cdot & s_{M-1} \\ s_1 & s_2 & \cdot & \cdot & \cdot \\ \cdot & \cdot & \cdot & \cdot & \cdot \\ \cdot & \cdot & \cdot & \cdot & \cdot \\ \cdot & \cdot & \cdot & \cdot & \cdot \\ s_{L-1} & \cdot & \cdot & \cdot & s_{N-1} \end{bmatrix} \tag{3}$$

The values of $L$ and $M$ should be chosen greater than the number of expected exponentially damped sinusoids $K$. The sum of $L$ and $M$ should be equal to the number of data points $N$ increased by one. It was proven that the method gave the best results when the relation was in the range $0.5 \leq L/M \leq 2.0$. The values outside that region may cause increase in the statistical error. It can be also noted that it is recommended to chose such parameters $L$ and $M$ to get matrix S as square as possible. In the

next step the data matrix $S$ is decomposed into a product of three matrices by application of SVD. (4)

$$S_{L \times M} = U_{L \times L} \Sigma_{L \times M} V_{M \times M}^H \tag{4}$$

According to Singular Value Decomposition, the data matrix $S$ is decomposed into a product of matrices. $U$ and $V$ are unitary matrices in which the columns are singular vectors and the symbol $H$ denotes Hermitian conjugation. $\Sigma$ is a diagonal matrix where entries on the main diagonal are the singular values. It should be mentioned that all singular values become nonzero when noise is present in the signal (5). On the other hand, the signal-noise-ratio of singular values related to the noise are smaller than the signal-related singular values. In the next step of the procedure the matrix $S$ should be truncated into a matrix $S_K$.

$$S_K = U_K \Sigma_K V_K^H \tag{5}$$

The symbol $K$, which is the number of sinusoids describing the measured signal, in that case corresponds to the number of columns chosen in the matrices $U$ and $V$. In the case of $\Sigma$ it denotes $K \times K$ upper-left submatrix. Now the task is to find a matrix that can transform one into another. That problem is solved by computation of the least square solution $E$ of the system. $V^{(t)}$ and $V^{(b)}$ are matrices derived from $V_K$ where the first and last row is omitted, respectively (6).

$$V_K^{(t)} E^H \approx V_K^{(b)} \tag{6}$$

When the equation is calculated, $K$ eigenvalues of $E$ give the signal pole estimates. Having those values it is easy to get the damping coefficient $d_k$ and frequencies $f_k$ (7).

$$z_n = a_k e^{\left(-d_k + j\,2\,\Pi f_k\right)n} \tag{7}$$

In this way estimates $z_k$ can be filled in the model equation and the least squares solution of $c_k$ computed (8).

$$c_k = a_k e^{j\Phi_k} \tag{8}$$

Finally, the amplitudes $a_k$ and phases $\Phi_k$ are obtained (9).

$$s_n \approx \sum_{k=1}^{K} c_k z_k^n \tag{9}$$

The most time consuming part of HSVD is the computation of the SVD of $L \times M$ matrix, whose time complexity is even of the 3rd order. The least square solution algorithm can be computed efficiently by applying correct methods. In this paper it can be noted that the full SVD is not required since only the first $K$ columns of matrices are necessary. Therefore improvements of HSVD are based on alternative matrix decomposition. Modification of the HSVD was introduced thanks to the Lanczos algorithm. The HLSVD computes only those singular values and vectors that represents the signal, ignoring all the others and exploiting the Hankel structure of the data matrix. By invoking The HLSVD the execution time of the SVD can be reduced. The algorithm has the following disadvantage, it can slow down in the case of a repeated or a close to singular value.

The described in the paper HSVD method requires in the first part of the algorithm nonlinear least squares which can be computed efficiently, while the SVD computation needs a lot of computational effort. In fact, in order to estimate the signal subspace only the calculated number of columns of the data matrix are required. Therefore instead of applying the full SVD to the data matrix, a modification of matrix decomposition, already present in the numerical linear algebra, can be used. Such a composition is called low-rank revealing decomposition. It computes only approximations of the desired signal subspace, resulting in significant time savings.

## 3. Preprocessing techniques

For that moment it was assumed that the given FID contains only recorded signal. However, after measurements, a spectrum will be rarely satisfactory in all respects. A common signal will be corrupted by the random noise with different degree, which can be identified by artifacts, the low signal to noise ratio, unwanted peak shapes or the limited resolution. Spectrum properties can be improved by multiplication of the spectrum with a more satisfactory lineshape function. In practice it means that the FID is multiplied by a time-domain filter, which represents the FT of the needed frequency-domain lineshape function. That process can be called windowing, apodization or simply filtering in the time domain. There are many approaches leading to filter FID- Dolph-Chebycheff window, Hamming filter, Kaiser filter, Hanning window, Savitzky-Golay filter or Lorentzian-to-Gaussian transformation. Filtering can give correction of the noise present in the signal, on the other hand it should be mentioned that each filtering method is applied for different reason. Additionally, filtering should be used carefully, because with an improper filter important information may be erased. According to equation (5) it is possible to neglect in the HSVD method the noise connected to less important single values (10) and in that case filtering is completed. Here the most important part consists in a correct choice of the number of components. Additionally, basing on that idea Cadzow filtering has been introduced [4].

$$H_{LxM} = \begin{bmatrix} U & U_0 \end{bmatrix}_{LxL} \begin{bmatrix} \Sigma & 0 \\ 0 & \Sigma_0 \end{bmatrix}_{LxM} \begin{bmatrix} V^H \\ V_0^H \end{bmatrix}_{MxM} \qquad (10)$$

The shape of peaks occurring in the spectrum after FT can be different than expected. The beginning of FID, more precisely the starting point of the sinusoidal FID in the time domain, is called a phase which is responsible for the shape of peaks. The phase error, which can disturb the final result, influences all possible angles, also the angle between absorptive and dispersive. In order to obtain the needed 0 degrees error (absorptive) after the FT, the spectrum has to be phase corrected. Raw MRS data (FID) consists of a real and imaginary part for each point in the time domain. After the FT the result has also a real and imaginary value for each frequency point. In an ideal case, when the phase is 0 degree, the real part of the spectrum becomes absorptive, which means that the shape of peaks is normal and the imaginary part stays dispersive, which corresponds to up/down mode. Normally it follows that the spectrum is mixed with absorptive and dispersive in various proportions. The phase correction is performed by calculating a linear combination of the real and imaginary parts of the spectrum. Mathematically it corresponds to the sum of two multiplications, the real spectrum with the cosine of the desired angle and the imaginary spectrum with the sine of the desired angle, which is in fact representation of rotation of two perpendicular vectors coming from the real and imaginary spectra. The main problem of phase correction focuses actually on finding a correct phase rotation angle, which is dependent on the chemical shift. It can be done manually but the challenge is to find that angle in an automatic way. In terms of the HSVD each metabolite is described by 4 parameters including a phase (11). In order to perform correction, it is obligatory to correct each phase value [5].

$$s_{corr} = s_k e^{j\Phi_{corr}} \qquad (11)$$

Spectrum lines often are not situated on a float line. In the ideal case it would be a straight, horizontal line representing zero intensity. Underline background in the spectrum is called a baseline and it comes from the average of the noise part of the spectrum. The main reason of rough baseline comes from hardware imperfections. Incorrect data are generated at the beginning of the FID, when a device is still recovering from the shock of the RF pulse. Hence the baseline signals can be recognized as a fast decaying component at the start of the FID. This becomes a problem in the spectrum analysis in the case of estimation of peak areas. The correction of the baseline consists in removal of the background signal in the spectrum. There are several approaches of that correction. A common method relies on finding several points in the spectrum, connecting them with a smooth line and subtract from the original spectrum (12).

$$FID_{corrected} = FID_{Total} - Baseline \qquad (12)$$

Very often the final result of the MRS is disturbed by phenomena called eddy currents, which are present during measurements due to the change of magnetic fields. Eddy currents induce additional time dependent magnetic fields which interact with the final result. Most MRS scanners already perform eddy current correction and it does not represent significant issue in the MRS, but there exist methods for elimination of that problem. The common practice in the MRS is to use the results of reference FID before and after suppression of water. Thanks to given data the signal free of eddy currents can be retrieved by the subtraction of the phase taken from the FID after suppression of water from the reference FID before suppression (13). Eddy currents influence mainly the amplitude of the FID, which may cause disturbances and artifacts in the spectrum [6].

$$s_{corr} = s_{unsuppressedwater} e^{j\Phi_{corr}} \qquad (13)$$

The MRS signal is not free of noises as it was mentioned before. Artifacts, which can be present in the signal are usually connected with the signal strength and therefore even small changes can influence the signal, finally disturbing the shapes and height of peaks. The possibility of removing a particular signal or a group of signals would be in that case a great advantage. Additionally, it could help in reduction of the complexity of the signal which is taken into consideration and as a result improve the accuracy of the parameters of the MRS. A good example of the unwanted compound in the MRS is a water signal, whose concentration in a human brain or other tissue is even several times higher than the concentration of different compounds and metabolites. For that reason the water signal is often removed from the final result during measurements. However, due to imperfection of current MRS scanners and specific nature of water, water suppression is never ideal and a significant part of water remains in the final spectrum, which causes the need to apply an additional post processing technique. The remaining water signal can cause a problem in the further processing analysis due to the complicated lineshape. A few approaches and strategies leading to removing the given signal from the FID can be applied by meaning of different properties of the signal. The first method which can be used is based on the fact that the frequency of water resonance is different than that of other compounds. Therefore correct filters can be applied to the signal, the water signal can be obtained and finally subtracted from the original FID. It should be noted that a filter has to be chosen carefully, otherwise improper filtering may have influence on the remaining signal. The filtering procedure has ability of removal of the water spectral line. The second possibility of water removal can be used partially during baseline correction. Alternatively the method of modeling can be used. Having parameters of each signal, the water component can be detected and removed from the original signal. Using the HSVD it is sufficient to detect metabolite of water and finally subtract it from the FID (14) [7].

$$FID_{corrected} = FID_{Total} - FID_{Water} \qquad (14)$$

## 4. Results

The analyzed data were taken from the Institute of Oncology in Gliwice and were tested by the HSVD algorithm prepared in Matlab. The data came from the phantom (having basic metabolites inside) analysis measured by a scanner GE, however for visualization in the paper one of the result was taken. In the first step the script should correctly read the data from GE files. The data are arranged in one file which consists of header and complex numbers related to water and other part of the signal. The Matlab script is able to show the final result in the form of a spectrum plot which is the real part of the Fourier Transform. Each step of the preprocessing method was performed separately. At first the data were modelled with the HSVD in terms of 32 components. All the other components were neglected and assumed to be the noise. The result of such filtering is shown in Fig. 1.

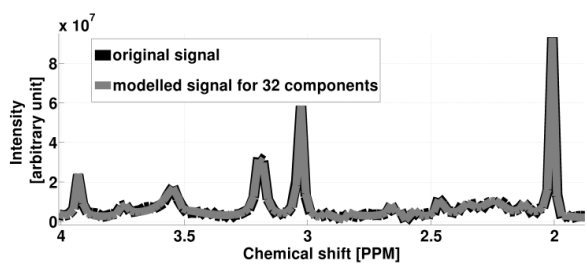

Fig. 1.    Filtering with the limited number of components
Rys. 1.    Filtrowanie z ograniczoną liczbą komponentów

In order to show the phase correction, the main component of the signal NAA (in 2.01 ppm) was distinguished. Thanks to the decomposition performed by the HSVD and in the next step the correct identification of the NAA, one can get all the parameters including the phase. For that case the value was corrected (however the correction was by a very small value) and the result is shown in Fig. 2.

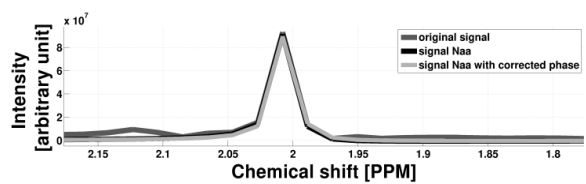

Fig. 2.    Phase correction shown on the NAA signal
Rys. 2.    Korekcja fazy pokazana na sygnale NAA

In the next step the water signal was modelled with the use of the HSVD and subtracted from the whole spectrum. Depending on the measurement, the remaining part of the water signal influences other part of spectrum. That impact is shown in Fig. 3. The baseline correction was skipped in the results due to the fact that the phantom data were not occupied by this problem.

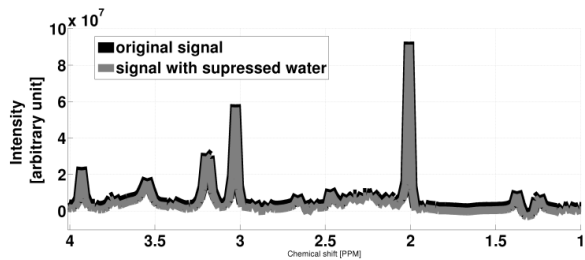

Fig. 3.    Influence of removal of the water signal on the whole spectrum
Rys. 3.    Wpływ usunięcia sygnału wody na całe widmo

Some scanners have eddy current correction implemented on a hardware side, however the GE has not. PThe pocedure of correction is described and the modelled result with the HSVD method is presented in Fig. 4.

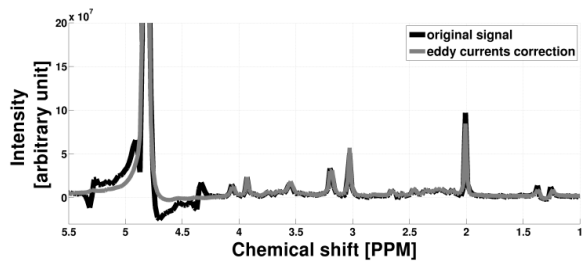

Fig. 4.    Eddy current correction
Rys. 4.    Korekcja prądów wirowych

## 5. Conclusions

In the paper the author has shown and described a few preprocessing steps of the magnetic resonance spectroscopy emphasizing what are the reasons of occurrence of particular problems and how to deal with them. The author has introduced the HSVD method which is commonly used in the MRS analysis for signal decomposition and concentration analysis. The author has proposed a development of that approach and applied the method to the preprocessing steps. The obtained results show that the method HSVD can be successfully used not only as a decomposition tool but also in the preprocessing. Preprocessing techniques are used depending on a situation and not each of them has to be used at the same time. The signal prepared in such a way may be used for the further analysis – concentration investigation.

*Work has been supported by BK-515/rau2/2013/9 under project „Development of methods correcting spectrum of magnetic resonance".*
*This work was partially supported by POIG.02.03.01-24-099/13 grant: „GCONiI - Upper-Silesian Center for Scientific Computation".*

## 6. References

[1]  Gunther H.: NMR SPECTROSCOPY Basic Principles, Concepts, and Applications in Chemistry. John Willey Sons, XIII-XIV, 1992.
[2]  Jiru F.: Introduction to post-processing techniques. European Journal of Radiology, vol. 67, pp. 202–217, 2008.
[3]  Pijnappel W. W. F., Van Den Boogaart A., de Beer R. and Van Ormondt D.: SVD-Based Quantification of Magnetic Resonance Signals. Journal of Magnetic Resonance, vol. 97, pp. 122-134, 1992.
[4]  Cadzow J. A.: Signal Enhancement – A Composite Property Mapping Algorithm. IEEE Transactions on Acoustic Speech and Signal Processing vol. 36, pp. 49 – 62, 1988.
[5]  Chen L., Weng Z., Goh L., Garland M.: An efficient algorithm for automatic phase correction of NMR spectra based on entropy minimization. Journal of Magnetic Resonance vol. 158, pp. 164–168, 2002.
[6]  Klose U.: In vivo proton spectroscopy in presence of eddy currents, Magn. Reson. Med., vol. 14, pp. 26–30, 1990.
[7]  Vanhamme L., Fierro R., Van Huffel S., de Beer R.: Fast removal of residual water in proton spectra. Journal of Magnetic Resonance, vol. 132(2), pp. 197-203, 1997.